\documentstyle[epsf,12pt]{article}

\newlength{\figsize}
\begin{document}

\begin{titlepage}

\begin{tabbing}
\` Oxford preprint OUTP-02-27P \\
\end{tabbing}
 
\vspace*{0.7in}
 
\begin{center}
{\large\bf SU(N) gauge theories in 2+1 dimensions\\
 -- further results \\}
\vspace*{0.5in}
{Biagio Lucini and Michael Teper\\
\vspace*{.2in}
Theoretical Physics, University of Oxford,\\
1 Keble Road, Oxford, OX1 3NP, U.K.\\
}
\end{center}

\vspace*{1.25in}

\begin{center}
{\bf Abstract}
\end{center}

We calculate the string tension and part of the mass spectrum
of SU(4) and SU(6) gauge theories in 2+1 dimensions using
lattice techniques. We combine these new results with 
older results for $N_c=2,...,5$ so as to obtain
more accurate extrapolations to $N_c = \infty$. The 
qualitative conclusions of the earlier work are unchanged:
SU($N_c$) theories in 2+1 dimensions are linearly confining 
as $N_c \to \infty$; the limit is achieved by keeping $g^2N_c$
fixed; SU(3), and even SU(2),
are `close' to SU($\infty$). We obtain more convincing 
evidence than before that the leading large-$N_c$ correction 
is $O(1/N^2_c)$. We look for the multiplication of states
that one expects in simple flux loop models of glueballs,
but find no evidence for this.

\end{titlepage}

\setcounter{page}{1}
\newpage
\pagestyle{plain}

\section{Introduction}
\label		{intro}

As a by-product of our recent work on $k$-strings in D=2+1
SU(4) and SU(6) gauge theories 
\cite{blmt-kstring}
we also calculated the lightest masses with 
$J^{PC} = 0^{++},0^{--},2^{++},2^{--}$ quantum numbers.
In this note we combine these new results
with the previous work of one of the authors
\cite{mt98}
so as to obtain more accurate information about the
$N_c = \infty$ limit and the $N_c$ dependence of
SU($N_c$) gauge theories in D=2+1.

This note is intended as an addendum and we refer
the reader to 
\cite{mt98}
as well as our related work in D=3+1
\cite{blmt-kstring,blmtd4}
for a discussion of the (well-known) physics motivation
\cite{largeN}
as well as the (mostly standard) lattice techniques.

In the next section we discuss the way our analysis
differs from that of 
\cite{mt98}
and we perform a partial re-analysis of that earlier
work so as to allow us to include it in a combined
analysis with the new calculations. We then turn
to an analysis of the string tension, $\sigma$,
which is the physical quantity we calculate most
accurately. We confirm that the $N_c=\infty$ limit is
achieved by keeping constant the 't Hooft coupling,
$g^2N_c$, and we are able to determine quite accurately
that the leading corrections are $O(1/N^2_c)$. We then
analyse the glueball masses and confirm that they also 
show little variation with $N_c$. Finally we describe
some calculations designed to reveal the increasing
multiplication of states with increasing $N_c$, 
which is expected
\cite{rjmt}
in models where glueballs are composed of closed loops
of flux. We find no evidence for this in our 
exploratory study.

\section{Some technical details}

Our lattices are cubic with periodic boundary conditions.
We use the standard plaquette action 
\begin{equation}
S = \beta \sum_{p}\{1-{1\over N_c}ReTr U_p\}
\label{B1}
\end{equation}
where the ordered product of the SU($N_c$) matrices around 
the boundary of the plaquette $p$ is represented by $U_p$.
In the continuum limit this becomes the usual
Yang-Mills action with
\begin{equation}
\beta = {{2N_c} \over {ag^2}}
\label{B2}
\end{equation}
where we recall that $g^2$ has dimensions of mass in D=2+1,
so that the dimensionless bare coupling, being a coupling on
the scale $a$, is just $ag^2$. The simulations are performed
with a combination of heat bath and over-relaxation updates, 
as described in 
\cite{mt98}.
Our SU(4) and SU(6) updates involve 6 and 15 SU(2) subgroups
respectively. Masses are calculated from the exponential
fall-off of appropriate correlators at large enough $t$,
again as described in
\cite{mt98}.
In Table 1 we list the values of $\beta$ at which we perform
calculations, together with the lattice size, $L$, and
the value of the average plaquette. In Tables 2 and 3
we list the masses we obtain for SU(4) and SU(6)
respectively.

Our present calculations differ from 
\cite{mt98}
in some technical details. Firstly, the mass calculations 
have been performed for fewer quantum numbers and for
a more limited range of operators. (The
primary purpose of these calculations was to obtain the
$k$-string tensions discussed in 
\cite{blmt-kstring}.)
Since the mass calculations involve a variational selection 
of the `best' glueball wavefunction from the basis of
operators used, there is a danger that this might introduce a
problem in combining the present results with those of
\cite{mt98}.
However we shall see that the new and old SU(4) calculations
agree with each other within errors which tells us that
there is in fact no reason to worry.

The second difference is that we now do exponential fits
to the correlation functions (as described in
\cite{blmtd4})
rather than relying on effective masses as in 
\cite{mt98}.
Again, the fact that the new and old SU(4) calculations
agree with each other within errors reassures us that
this introduces no significant relative bias.

Finally we use standard Gaussian error estimates rather 
than the somewhat idiosyncratic algorithm described and 
used in
\cite{mt98}.
In order to be able to perform a unified statistical
analysis of the old and new calculations, we have
redone all the (relevant) continuum extrapolations in
\cite{mt98},
obtaining the values shown in Table 4. We note that
the new errors are typically about 30\% smaller than
those presented in 
\cite{mt98}.
\section{The string tension}

Our first step is to calculate the mass, $m_P(L)$, of a flux tube 
that winds around our $L\times L$ spatial torus, as described in 
\cite{mt98,blmt-kstring}.
We then extract the string tension, $\sigma$, using
\begin{equation}
a m_P(L) = a^2\sigma L - {\pi\over{6L}}.
\label{C1}
\end{equation}
This relation assumes that we have linear confinement and that
the leading correction is that of a simple periodic bosonic string
\cite{stringcor}. 
In the case of SU(2) there is now rather convincing evidence 
\cite{blmt-kstring}
for both these assumptions. As an example of what one can do for
larger groups, such as SU(4), we plot $am_P(L)$ as a 
function of $L$ at $\beta=28$ in Fig.1. (The masses on the
smaller volumes are listed in Table 11 of
\cite{blmt-kstring}.)
We see the approximate linear rise confirming the persistence
of linear confinement for $N_c=4$. (Note that the largest 
string length is $16a \simeq 4/\surd\sigma$ which should
be long enough to realistically address this issue.)
We also see that the
bosonic string correction fills in most of the deviation
from linearity at larger $L$. We shall therefore use
eqn(\ref{C1}) to extract the string tension from the 
masses of flux loops that are longer than 
$\simeq 3/\surd\sigma$, where one typically finds
\cite{blmt-kstring}
that eqn(\ref{C1}) works well.

Having calculated the string tension as a function of $\beta$, 
we follow 
\cite{mt98}
and extrapolate to the continuum limit using
\begin{equation}
\beta_I a \surd \sigma = c_0 + {{c_1}\over\beta_I}
\label{C2}
\end{equation}
where $\beta_I$ is the mean field improved coupling
\begin{equation}
\beta_I  =  \beta \times  \langle {1\over{N_c}} Tr U_p \rangle.
\label{C3}
\end{equation}
We obtain the continuum string tension from the fitted value
of $ c_0=  2 N_c {{\surd\sigma}/{g^2}}$. The values of
${{\surd\sigma}/{g^2}}$ thus
obtained are listed in Table~\ref{table_Kcont}.

In Fig.\ref{fig_stringN} we plot the calculated values of
$\surd\sigma/g^2$ against $N_c$. We immediately observe that 
the variation approaches a linear form for larger $N_c$
\begin{equation}
{ {\surd\sigma} \over {g^2} } \propto N_c
\label{C8}
\end{equation}
and indeed is nearly linear even down to $N_c=2$. This
confirms the usual diagrammatic expectation
\cite{largeN}
that a smooth large-$N_c$ limit is obtained by keeping
constant the 't Hooft coupling $\lambda\equiv g^2N_c$.

We can also ask what is the power of the leading correction.
To do so we fit our string tensions with the functional form
\begin{equation}
{ {\surd\sigma} \over {g^2 N_c} } = 
c_0 + {{c_1} \over {N_c^{\alpha}}}.
\label{C4}
\end{equation}
In Fig.\ref{fig_corrN} we show how the goodness of fit varies
with the power $\alpha$. From this we can infer that 
\begin{equation}
\alpha = 1.97 \pm 0.21
\label{C5}
\end{equation}
if we choose errors corresponding to a 20\% confidence level.
If we assume that 
the power should be an integer, then only one value is
allowed: $\alpha=2$. This is in agreement with the 
usual diagrammatic expectations
\cite{largeN}.

Fitting our calculated values with $\alpha=2$, we obtain
\begin{equation}
{ {\surd\sigma} \over {g^2 N_c} } = 
0.19755(34) - {{0.1200(29)} \over {N_c^2}}
\label{C12}
\end{equation}
This fit has a good confidence level, $\sim 70 \%$. 
Thus, despite our smaller errors and extended range in $N_c$,
the observation in
\cite{mt98}
still stands:  we can describe
the string tension of $SU(N_c)$ gauge theories, all the way 
down to $SU(2)$, by that of the $SU(\infty)$ theory supplemented
by the leading correction with a modest coefficient.

\section{The mass spectrum}

We list in Tables~\ref{table_msu4} and ~\ref{table_msu6} 
our calculated masses. We know 
\cite{Sym}
that with the plaquette action the leading  lattice correction to 
dimensionless ratios of physical quantities, such as $m_G/\surd\sigma$ 
is $O(a^2)$. Thus we extrapolate to the continuum limit using
\begin{equation}
 {{m_G(a)} \over {\surd\sigma(a)}}
 = 
 {{m_G(a=0)} \over {\surd\sigma(a=0)}} + ca^2\sigma.
\label{D1}
\end{equation}
The results are shown in Table~\ref{table_mcont}. All these
extrapolations have a good $\chi^2$ apart from those indicated 
by an asterisk. We also show similar extrapolations for
the masses calculated in   
\cite{mt98}
but with the errors recalculated as described in Section 2.

We first note that the old and new SU(4) calculations agree, 
within errors, for the lightest, most accurately determined
states. This reassures us that the different
procedures used in these calculations do not lead to a 
relative systematic difference
that would undermine a combined statistical analysis.

The variation of our mass ratios with $N_c$ is weak 
which encourages us to try to fit with just the leading
large-$N_c$ correction:
\begin{equation}
 {{m_G} \over {g^2N_c}}
 = 
 d_0 + {{d_1} \over {N_c^2}}
\label{D4}
\end{equation}
where $d_0 = \lim_{N_c\to\infty} m_G/g^2N_c$.
The results of such fits are listed in Table~\ref{table_massgN}
where we also produce corresponding values of $m_G/\surd\sigma$,
which have been obtained by using eqn(\ref{C12}). 
We note that we obtain acceptable fits in all cases, 
except in the case of the rather heavy $2^{++}$ where the low 
value in SU(6) makes it impossible to find a good fit.

To illustrate all this we plot $m_G/g^2N_c$ against
$1/N_c^2$ for the $0^{++}$ and $0^{--}$ glueballs in 
Fig.\ref{fig_msuN}. We observe that the dependence of our 
masses on $N_c$ is really very weak over the whole range of
$N_c$.

\section{Are glueballs loops of flux?}

If SU(3) glueballs are composed of closed loops of
fundamental (k=1) flux then in  SU($N_c \geq 4$) there will 
be extra states composed of closed loops of $k\not= 1$ flux 
\cite{rjmt}.
In simple, but natural, string models of glueballs
\cite{rjmt,ip}
the masses of such states will be scaled up by a factor 
of approximately $\sqrt{\sigma_k/\sigma}$,  where 
$\sigma_k$ is the $k$-string tension (so that
$\sigma_{k=1}\equiv \sigma$). These string tensions
have been recently calculated and satisfy Casimir scaling,
$\sigma_k/\sigma \simeq k(N_c-k)/(N_c-1)$,
to a good approximation
\cite{blmt-kstring}.
Is there any sign of such extra states in our calculations?

We recall that our glueball operators are composed of products
of highly `smeared' SU($N_c$) matrices around the boundaries 
of squares and rectangles
\cite{mt98}.
When we take products of these matrices around the torus we
obtain a basis of operators that typically provides 
a good wavefunctional for the winding fundamental flux tube and
so we may regard our usual glueball operators as being closed 
loops of fundamental flux. To construct corresponding loops
of $k=2$ strings we take the fundamental loop, $l$, twice around 
the boundary in the form ${\mathrm Tr}(l.l)$ or ${\mathrm Tr}l$. 
We can project onto the antisymmetric $k=2$  representation,
$k=2A$, as well. We look for extra states in the $k=2$ and 
$k=2A$ bases of operators as compared to the $k=1$ basis.
We note that the lightest few states obtained in the
usual $k=1$ basis seem to continue well from $N_c=2,3$ to
larger $N_c$
\cite{mt98}
and so seem to possess no extra states at larger $N_c$.

We calculate the spectrum of states separately using as a basis
either the usual $k=1$ operators, or the $k=2$ operators or
the $k=2A$ operators. We do this for SU(6) where mixings and
decays, which might serve to obscure the effects we are searching
for, should be weak. In SU(6) any such new states should have 
masses that are larger by a factor $\sim \sqrt{1.6} \simeq 1.26$.
In Table~\ref{table_iptest} we show the $0^{++}$ states
obtained using these different bases of operators. Masses are
obtained from exponential fits. For heavier states, indicated
by a single asterisk, we take the effective mass from 
$t=a$ to $t=2a$. for the very heaviest states, indicated
by a double asterisk, we take the effective mass from 
$t=0$ to $t=a$. What we see is that there is no clear sign of
extra states in the $k=2A$ sector as compared to the
$k=1$ sector, in the expected mass range. It is interesting
to note that the one clear extra state in the $k=2$ sector,
with a mass of $am \simeq 1.20$, fits in well with being
just the scattering state composed of two ground state
$0^{++}$ glueballs in an $S$-wave. (Large-$N_c$ counting
will suppress its appearance in the $k=1$ sector.) We
have performed similar comparisons in the $0^{--}$ and
$2^{++}$ sectors with similar conclusions.

Although our results look quite negative, they should
be regarded as preliminary. Our operators are based
on closed loops whose sizes are not much larger than the
transverse extent of the smeared matrix, and which
might therefore not be a good basis for closed 
`string-like' states. So calculations with larger loops
need to be performed. Moreover the lighter states may
be smaller and may be the least string-like.
Thus one needs an accurate comparison for higher mass
states, which is something we would not claim to have achieved 
in the present calculation.

\section{Conclusions}

In this paper we have confirmed the main conclusions of 
\cite{mt98}
over a larger range of $N_c$. We have also looked
for an increase in the density of states of the kind
that naturally arises 
\cite{rjmt,ip}
in models where  glueballs are composed of closed loops 
of flux. In what is admittedly a crude study, we found no
sign of such extra states.

One motivation for these large-$N_c$ calculations is to
provide theorists who are attempting to develop (mainly)
analytic methods, with something to compare against.
Thus we find it interesting to note that
the string tension derived as a first approximation in
\cite{nair}
\begin{equation}
 {{\sqrt\sigma} \over {g^2}}
 = 
 \sqrt{{N_c^2 -1} \over {8\pi}}
\label{F1}
\end{equation}
is in fact extremely close to all our calculated values.
As for the glueball spectrum, we note the predictions in 
\cite{dalley}
for the lightest glueball masses in the $N_c\to\infty$ 
limit: $m_{0^{++}} = 4.10(13) \sqrt\sigma$, 
$m_{0^{--}}/m_{0^{++}}=1.35(5)$ and
$m_{2^{++}}/m_{0^{++}}=1.60(17)$. The first and last
of these predictions are consistent with our values, 
while the second is only two standard deviations out. 
We also note that recent AdS/CFT calculations of glueball
masses (see
\cite{terning}
for a review) are compatible with our values.
Finally one should add that insights into the
significance of the detailed mass spectrum may
be obtained through a comparison with related
D=2+1 field theories
\cite{caselle}.

\section*{Acknowledgments}
These calculations were carried out on Alpha Compaq workstations 
in Oxford Theoretical Physics, funded by PPARC and EPSRC grants.
One of us (BL) was  supported by a PPARC and then by a
EU Marie Sklodowska-Curie postdoctoral fellowship.

\newpage

\begin{table}
\begin{center}
\begin{tabular}{|c|c|c||c|c|c|}\hline
\multicolumn{3}{|c||}{SU(4)} & \multicolumn{3}{c|}{SU(6)} \\ \hline
$\beta$ & $L$ & plaq & $\beta$ & $L$ & plaq  \\ \hline
60.0 & 32 & 0.914426(2)  & 108.0 & 24 & 0.887953(2)   \\
45.0 & 24 & 0.884724(3)  & 75.0  & 16 & 0.835392(6)   \\ 
33.0 & 16 & 0.840190(7)  & 60.0  & 12 & 0.790321(8)   \\ 
28.0 & 16 & 0.809332(6)  & 49.0  & 10 & 0.736809(12)  \\
28.0 & 12 & 0.809350(15) & 42.0  & 8  & 0.684495(20)  \\
21.0 & 10 & 0.737640(15) & & & \\
18.0 & 8  & 0.68593(4)   & & & \\ \hline
\end{tabular}
\caption{\label{table_plaqsu46}
Average SU(4) and SU(6) plaquette values for the new calculations.}
\end{center}
\end{table}

\begin{table}
\begin{center}
\begin{tabular}{|l||l|l|l|l|l|l|l|}\hline
state & $\beta=18$ & $\beta=21$ & $\beta=28$ & $\beta=28$ & $\beta=33$ & $\beta=45$ &
$\beta=60$   \\
      & L=8  & L=10  & L=12  & L=16  & L=16  &  L=24   &  L=32   \\ \hline
$l$     & 1.50(2) & 1.223(11) & 0.715(4) & 0.9937(58) & 0.6622(29) & 0.4974(22) & 0.3571(14) \\ \hline 
$0^{++}$  & 1.85(6) & 1.54(4) & 1.057(15) & 1.082(11) & 0.872(10) & 0.620(5) & 0.4592(27) \\
$0^{++\ast}$  & 2.5(3) & 2.29(15) & 1.53(3) & 1.56(4) & 1.289(23) & 0.940(8) & 0.682(4) \\
$0^{++\ast\ast}$ & -- & 2.58(34) & 1.77(5) & 1.88(6) & 1.506(24) & 1.117(33) & 0.842(5) \\
$0^{--}$         & 2.7(5) & 2.11(14) & 1.61(3) & 1.53(5) & 1.306(22) & 0.895(20) & 0.6736(34) \\
$0^{--\ast}$     & -- & 2.16(26) & 1.97(9) & 1.96(14) & 1.65(4) & 1.233(16) & 0.852(14) \\
$2^{++}$         & -- & 2.4(3) & 1.5(2) & 1.81(5) & 1.502(26) & 1.051(12) & 0.773(6) \\
$2^{++\ast}$     & -- & -- & 2.35(16) & 2.23(11) & 1.71(6) & 1.22(6) & 0.887(18) \\
$2^{--}$         & -- & -- & 2.03(11) & 2.21(22) & 1.81(5) & 1.228(13) & 0.921(10) \\ \hline
\end{tabular}
\caption{\label{table_msu4}
Masses of glueballs, labelled by $J^{PC}$, and of the 
periodic flux loop, $l$, of length $L$ : in SU(4) at the
indicated values of $\beta$ on lattice volumes $L^3$.}
\end{center}
\end{table}

\begin{table}
\begin{center}
\begin{tabular}{|l||l|l|l|l|l|}\hline
state & $\beta=42$ & $\beta=49$ & $\beta=60$ & $\beta=75$ &
$\beta=108$   \\
      & L=8  & L=10  & L=12  & L=16  &  L=24   \\ \hline
$l$     & 1.453(14) & 1.194(9) & 0.8575(47) & 0.6825(31) & 0.4552(13) \\ \hline 
$0^{++}$         & 1.754(54) & 1.491(31) & 1.161(16) & 0.8885(42) & 0.5925(26) \\
$0^{++\ast}$     & 2.7(7) & 2.13(13) & 1.752(60) & 1.347(18) & 0.8769(57) \\
$0^{++\ast\ast}$ &  -- & 2.58(34) & 1.97(9) & 1.605(27) & 1.1226(83) \\
$0^{--}$         & 2.50(27) & 2.20(13) & 1.707(43) & 1.284(17) & 0.861(8) \\ 
$0^{--\ast}$     & -- & 2.8(4) & 2.14(11) & 1.66(4) & 1.119(12) \\ 
$2^{++}$         & 2.6(4) & 2.6(4) & 1.92(7) & 1.478(23) & 0.953(16) \\
$2^{++\ast}$     & -- & -- & 2.40(16) & 1.82(6) & 1.242(11) \\
$2^{--}$         & -- & -- &   2.3(2) & 1.79(5) & 1.197(10) \\ \hline 
\end{tabular}
\caption{\label{table_msu6}
Masses of glueballs, labelled by $J^{PC}$, and of the 
periodic flux loop, $l$, of length $L$ : in SU(6) at the
indicated values of $\beta$ on lattice volumes $L^3$.}
\end{center}
\end{table}

\begin{table}
\begin{center}
\begin{tabular}{|l||l|l|l|l||l|l|}\hline
\multicolumn{7}{|c|}{$m_G/\surd\sigma$} \\ \hline
state & SU(2) & SU(3) & SU(4) & SU(5)& SU(4) & SU(6) \\ \hline
$0^{++}$         & 4.716(21) & 4.330(24) & 4.239(34) & 4.180(39) & 4.235(25) & 4.196(27) \\
$0^{++\ast}$     & 6.78(7)   & 6.485(55) & 6.383(77) & 6.22(8) & 6.376(45) & 6.20(7) \\
$0^{++\ast\ast}$ & 8.07(10)  & 8.21(10) & 8.12(13) & 7.87(18) & 7.93(7) & 8.22(12) \\ 
$0^{--}$         &           & 6.464(48) & 6.27(6) & 6.06(11) & 6.230(44) & 6.097(80) \\ 
$0^{--\ast}$     &           & 8.14(8)  & 7.84(13) & 7.85(15) & 8.20(15)$^\ast$ & 7.98(15) \\ 
$2^{++}$         & 7.81(6)  & 7.12(7) & 7.14(8) & 7.15(12) & 7.17(8) & 6.67(18) \\
$2^{++\ast}$     &           &          & 8.50(17) & 8.56(15) & 8.06(22) & 8.89(20) \\
$2^{--}$         &           & 8.73(10) & 8.25(21) & 8.25(18) & 8.49(13) & 8.52(20) \\ \hline
\end{tabular}
\caption{\label{table_mcont}
Glueball masses in units of the string tension, in the continuum
limit. Reanalysis of \cite{mt98} on left; new calculations
on right.}
\end{center}
\end{table}

\begin{table}
\begin{center}
\begin{tabular}{|c|c|c||c|}\hline
group & $c_0$ & $c_1$ & $\surd\sigma/g^2$  \\ \hline
SU(4) &  6.072(16)  & -8.19(45)   & 0.7590(20) \\
SU(6) & 14.000(38)  & -46.5(2.6)  & 1.1667(32) \\ \hline
SU(2) &  1.3405(31) & -0.417(24)  & 0.3351(8)  \\
SU(3) &  3.3182(61) & -2.43(11)   & 0.5530(10) \\
SU(4) &  6.065(21)  & -7.74(73)   & 0.7581(26) \\
SU(5) &  9.657(38)  & -21.4(1.9)  & 0.9657(38) \\ \hline
\end{tabular}
\caption{\label{table_Kcont}
Continuum extrapolations of 
$\beta_I a\surd\sigma \to 2N_c \surd\sigma/g^2$ 
as in eqn(\ref{C1}). New calculations (top)
and reanalysed old calculations (bottom).}
\end{center}
\end{table}

\begin{table}
\begin{center}
\begin{tabular}{|l||c|l|c|l||c|}\hline
state & $\lim_{N_c\to\infty} m/g^2N_c$ & slope & $N_c\geq$ 
& $\chi^2/n_{df}$ & $\lim_{N_c\to\infty} m/\surd\sigma$ \\ \hline \hline
$0^{++}$         & 0.8116(36) & -0.090(28) & 2 & 0.9 & 4.108(20) \\ 
$0^{++\ast}$     & 1.227(9)   & -0.343(82) & 2 & 0.8 & 6.211(46) \\
$0^{++\ast\ast}$ & 1.65(4)    & -2.2(7)    & 4 & 1.6 & 8.35(20)  \\ 
$0^{--}$         & 1.176(14)  &  0.14(20)  & 3 & 0.2 & 5.953(71) \\
$0^{--\ast}$     & 1.535(28)  & -0.35(35)  & 3 & 1.0 & 7.77(14)  \\
$2^{++}$         & 1.359(12)  & -0.22(8)   & 2 & 2.9 & 6.88(6)   \\
$2^{++\ast}$     & 1.822(62)  & -3.9(1.3)  & 4 & 0.4 & 9.22(32)  \\
$2^{--}$         & 1.615(33)  & -0.10(42)  & 3 & 1.0 & 8.18(17)  \\ \hline
\end{tabular}
\caption{\label{table_massgN}
The large $N_c$ limit of the mass spectrum in units of 
$g^2N_c$; with the slope of the linear fit 
when plotted against $1/N_c^2$. Also the 
range of colours fitted and the $\chi^2$ per degree of 
freedom of the best fit. Last column uses the value of
$\lim_{N_c\to\infty}\surd\sigma/g^2N_c$ in eqn(\ref{C12}).}
\end{center}
\end{table}

\begin{table}
\begin{center}
\begin{tabular}{|c|c|c|}\hline
\multicolumn{3}{|c|}{$a m_{eff}$ ; $J^{PC}=0^{++}$} \\ \hline
 $k=1$ ; 10 ops &  $k=2A$ ; 10 ops & $k=2$  ; 20 ops \\ \hline
  0.592(4) &  0.594(4) &  0.593(4) \\
  0.88(1) &  0.90(1) &  0.89(1) \\
  1.14(1) &  1.17(1) &  1.13(1) \\
          &          &  1.20(1) \\
  1.30(2) &  1.29(6) &  1.31(2) \\
  1.47(2)$^\ast$ &  1.42(2)$^\ast$  &  1.46(2)$^\ast$  \\
  1.53(2)$^\ast$ &                  &  1.49(2)$^\ast$  \\
  1.78(4)$^\ast$ &  1.76(4)$^\ast$  &  1.72(4)$^\ast$ \\
   &   &  1.73(4)$^\ast$ \\
   &   &  1.65(3)$^\ast$ \\
   &   &  1.72(4)$^\ast$ \\
   2.9(2)$^{\ast\ast}$& 2.9(2)$^{\ast\ast}$  &  2.3(1)$^\ast$  \\ \hline
\end{tabular}
\caption{\label{table_iptest}
Spectrum of scalar masses at $\beta=108$ in SU(6) obtained
using operators based on fundamental flux loops ($k=1$)
doubly charged loops ($k=2$) and from the antisymmetric
combination of the latter ($k=2A$). States are listed in
the order of effective masses from $t=0$ to $t=a$ and are 
matched to guide the eye.}
\end{center}
\end{table}

\clearpage

\begin	{figure}[p]
\begin	{center}
\leavevmode
\setlength{\unitlength}{0.240900pt}
\ifx\plotpoint\undefined\newsavebox{\plotpoint}\fi
\sbox{\plotpoint}{\rule[-0.200pt]{0.400pt}{0.400pt}}%
\begin{picture}(1500,1800)(0,0)
\font\gnuplot=cmr10 at 12pt
\gnuplot
\sbox{\plotpoint}{\rule[-0.200pt]{0.400pt}{0.400pt}}%
\put(400.0,250.0){\rule[-0.200pt]{4.818pt}{0.400pt}}
\put(375,250){\makebox(0,0)[r]{\ \ {$0$}}}
\put(1405.0,250.0){\rule[-0.200pt]{4.818pt}{0.400pt}}
\put(400.0,500.0){\rule[-0.200pt]{4.818pt}{0.400pt}}
\put(375,500){\makebox(0,0)[r]{\ \ {$0.2$}}}
\put(1405.0,500.0){\rule[-0.200pt]{4.818pt}{0.400pt}}
\put(400.0,750.0){\rule[-0.200pt]{4.818pt}{0.400pt}}
\put(375,750){\makebox(0,0)[r]{\ \ {$0.4$}}}
\put(1405.0,750.0){\rule[-0.200pt]{4.818pt}{0.400pt}}
\put(400.0,1000.0){\rule[-0.200pt]{4.818pt}{0.400pt}}
\put(375,1000){\makebox(0,0)[r]{\ \ {$0.6$}}}
\put(1405.0,1000.0){\rule[-0.200pt]{4.818pt}{0.400pt}}
\put(400.0,1250.0){\rule[-0.200pt]{4.818pt}{0.400pt}}
\put(375,1250){\makebox(0,0)[r]{\ \ {$0.8$}}}
\put(1405.0,1250.0){\rule[-0.200pt]{4.818pt}{0.400pt}}
\put(400.0,1500.0){\rule[-0.200pt]{4.818pt}{0.400pt}}
\put(375,1500){\makebox(0,0)[r]{\ \ {$1$}}}
\put(1405.0,1500.0){\rule[-0.200pt]{4.818pt}{0.400pt}}
\put(605.0,250.0){\rule[-0.200pt]{0.400pt}{4.818pt}}
\put(605,200){\makebox(0,0){\ {$4$}}}
\put(605.0,1730.0){\rule[-0.200pt]{0.400pt}{4.818pt}}
\put(810.0,250.0){\rule[-0.200pt]{0.400pt}{4.818pt}}
\put(810,200){\makebox(0,0){\ {$8$}}}
\put(810.0,1730.0){\rule[-0.200pt]{0.400pt}{4.818pt}}
\put(1015.0,250.0){\rule[-0.200pt]{0.400pt}{4.818pt}}
\put(1015,200){\makebox(0,0){\ {$12$}}}
\put(1015.0,1730.0){\rule[-0.200pt]{0.400pt}{4.818pt}}
\put(1220.0,250.0){\rule[-0.200pt]{0.400pt}{4.818pt}}
\put(1220,200){\makebox(0,0){\ {$16$}}}
\put(1220.0,1730.0){\rule[-0.200pt]{0.400pt}{4.818pt}}
\put(400.0,250.0){\rule[-0.200pt]{246.922pt}{0.400pt}}
\put(1425.0,250.0){\rule[-0.200pt]{0.400pt}{361.350pt}}
\put(400.0,1750.0){\rule[-0.200pt]{246.922pt}{0.400pt}}
\put(50,1350){\makebox(0,0){\large{$am_l(L)$}}}
\put(912,25){\makebox(0,0){\large{$L$}}}
\put(400.0,250.0){\rule[-0.200pt]{0.400pt}{361.350pt}}
\put(605.0,417.0){\rule[-0.200pt]{0.400pt}{0.723pt}}
\put(595.0,417.0){\rule[-0.200pt]{4.818pt}{0.400pt}}
\put(595.0,420.0){\rule[-0.200pt]{4.818pt}{0.400pt}}
\put(708.0,589.0){\rule[-0.200pt]{0.400pt}{0.723pt}}
\put(698.0,589.0){\rule[-0.200pt]{4.818pt}{0.400pt}}
\put(698.0,592.0){\rule[-0.200pt]{4.818pt}{0.400pt}}
\put(810.0,780.0){\rule[-0.200pt]{0.400pt}{2.168pt}}
\put(800.0,780.0){\rule[-0.200pt]{4.818pt}{0.400pt}}
\put(800.0,789.0){\rule[-0.200pt]{4.818pt}{0.400pt}}
\put(913.0,961.0){\rule[-0.200pt]{0.400pt}{1.927pt}}
\put(903.0,961.0){\rule[-0.200pt]{4.818pt}{0.400pt}}
\put(903.0,969.0){\rule[-0.200pt]{4.818pt}{0.400pt}}
\put(1015.0,1139.0){\rule[-0.200pt]{0.400pt}{2.409pt}}
\put(1005.0,1139.0){\rule[-0.200pt]{4.818pt}{0.400pt}}
\put(1005.0,1149.0){\rule[-0.200pt]{4.818pt}{0.400pt}}
\put(1220.0,1485.0){\rule[-0.200pt]{0.400pt}{3.373pt}}
\put(1210.0,1485.0){\rule[-0.200pt]{4.818pt}{0.400pt}}
\put(605,418){\circle*{12}}
\put(708,591){\circle*{12}}
\put(810,785){\circle*{12}}
\put(913,965){\circle*{12}}
\put(1015,1144){\circle*{12}}
\put(1220,1492){\circle*{12}}
\put(1210.0,1499.0){\rule[-0.200pt]{4.818pt}{0.400pt}}
\sbox{\plotpoint}{\rule[-0.500pt]{1.000pt}{1.000pt}}%
\multiput(547,250)(6.104,19.838){2}{\usebox{\plotpoint}}
\put(560.07,289.38){\usebox{\plotpoint}}
\multiput(566,305)(6.981,19.546){2}{\usebox{\plotpoint}}
\put(581.50,347.84){\usebox{\plotpoint}}
\put(589.14,367.13){\usebox{\plotpoint}}
\multiput(597,385)(7.708,19.271){2}{\usebox{\plotpoint}}
\put(613.08,424.60){\usebox{\plotpoint}}
\put(621.56,443.54){\usebox{\plotpoint}}
\put(630.41,462.31){\usebox{\plotpoint}}
\multiput(638,479)(8.589,18.895){2}{\usebox{\plotpoint}}
\put(656.84,518.68){\usebox{\plotpoint}}
\put(665.85,537.38){\usebox{\plotpoint}}
\put(675.23,555.89){\usebox{\plotpoint}}
\put(684.68,574.37){\usebox{\plotpoint}}
\put(693.96,592.93){\usebox{\plotpoint}}
\put(703.50,611.36){\usebox{\plotpoint}}
\put(713.32,629.64){\usebox{\plotpoint}}
\put(722.67,648.17){\usebox{\plotpoint}}
\put(732.44,666.48){\usebox{\plotpoint}}
\put(742.78,684.48){\usebox{\plotpoint}}
\put(752.45,702.85){\usebox{\plotpoint}}
\multiput(762,721)(10.399,17.962){2}{\usebox{\plotpoint}}
\put(782.61,757.31){\usebox{\plotpoint}}
\put(792.30,775.66){\usebox{\plotpoint}}
\put(803.04,793.42){\usebox{\plotpoint}}
\put(813.18,811.53){\usebox{\plotpoint}}
\put(823.26,829.67){\usebox{\plotpoint}}
\put(834.03,847.41){\usebox{\plotpoint}}
\put(844.18,865.52){\usebox{\plotpoint}}
\put(854.94,883.26){\usebox{\plotpoint}}
\put(865.09,901.36){\usebox{\plotpoint}}
\put(875.57,919.28){\usebox{\plotpoint}}
\put(886.39,936.99){\usebox{\plotpoint}}
\put(896.93,954.87){\usebox{\plotpoint}}
\put(907.01,973.02){\usebox{\plotpoint}}
\put(918.25,990.46){\usebox{\plotpoint}}
\put(928.35,1008.59){\usebox{\plotpoint}}
\put(938.94,1026.45){\usebox{\plotpoint}}
\put(950.13,1043.92){\usebox{\plotpoint}}
\put(960.58,1061.85){\usebox{\plotpoint}}
\put(970.86,1079.88){\usebox{\plotpoint}}
\put(981.99,1097.39){\usebox{\plotpoint}}
\put(992.70,1115.17){\usebox{\plotpoint}}
\put(1003.77,1132.72){\usebox{\plotpoint}}
\put(1014.30,1150.61){\usebox{\plotpoint}}
\put(1025.09,1168.33){\usebox{\plotpoint}}
\put(1036.08,1185.93){\usebox{\plotpoint}}
\put(1046.60,1203.82){\usebox{\plotpoint}}
\put(1057.49,1221.49){\usebox{\plotpoint}}
\put(1068.38,1239.15){\usebox{\plotpoint}}
\put(1078.91,1257.04){\usebox{\plotpoint}}
\put(1090.18,1274.45){\usebox{\plotpoint}}
\put(1101.11,1292.08){\usebox{\plotpoint}}
\put(1111.63,1309.97){\usebox{\plotpoint}}
\put(1122.74,1327.50){\usebox{\plotpoint}}
\put(1133.41,1345.30){\usebox{\plotpoint}}
\put(1144.34,1362.94){\usebox{\plotpoint}}
\put(1155.60,1380.38){\usebox{\plotpoint}}
\put(1166.17,1398.24){\usebox{\plotpoint}}
\put(1177.83,1415.41){\usebox{\plotpoint}}
\put(1188.35,1433.30){\usebox{\plotpoint}}
\put(1199.10,1451.05){\usebox{\plotpoint}}
\put(1210.55,1468.34){\usebox{\plotpoint}}
\put(1221.08,1486.23){\usebox{\plotpoint}}
\put(1232.02,1503.85){\usebox{\plotpoint}}
\put(1243.28,1521.28){\usebox{\plotpoint}}
\put(1254.02,1539.03){\usebox{\plotpoint}}
\put(1265.17,1556.54){\usebox{\plotpoint}}
\put(1276.29,1574.06){\usebox{\plotpoint}}
\put(1286.97,1591.86){\usebox{\plotpoint}}
\put(1298.03,1609.42){\usebox{\plotpoint}}
\put(1309.11,1626.97){\usebox{\plotpoint}}
\put(1319.71,1644.81){\usebox{\plotpoint}}
\put(1331.32,1662.01){\usebox{\plotpoint}}
\put(1341.91,1679.85){\usebox{\plotpoint}}
\put(1353.59,1697.01){\usebox{\plotpoint}}
\put(1364.17,1714.87){\usebox{\plotpoint}}
\put(1375.32,1732.37){\usebox{\plotpoint}}
\put(1386,1750){\usebox{\plotpoint}}
\end{picture}

\end	{center}
\vskip 0.15in
\caption{The mass of a periodic flux loop of length $aL$ 
at $\beta=28$ in SU(4). The line shows a fit of the form
in eqn(\ref{C1}).}
\label{fig_string4}
\end 	{figure}
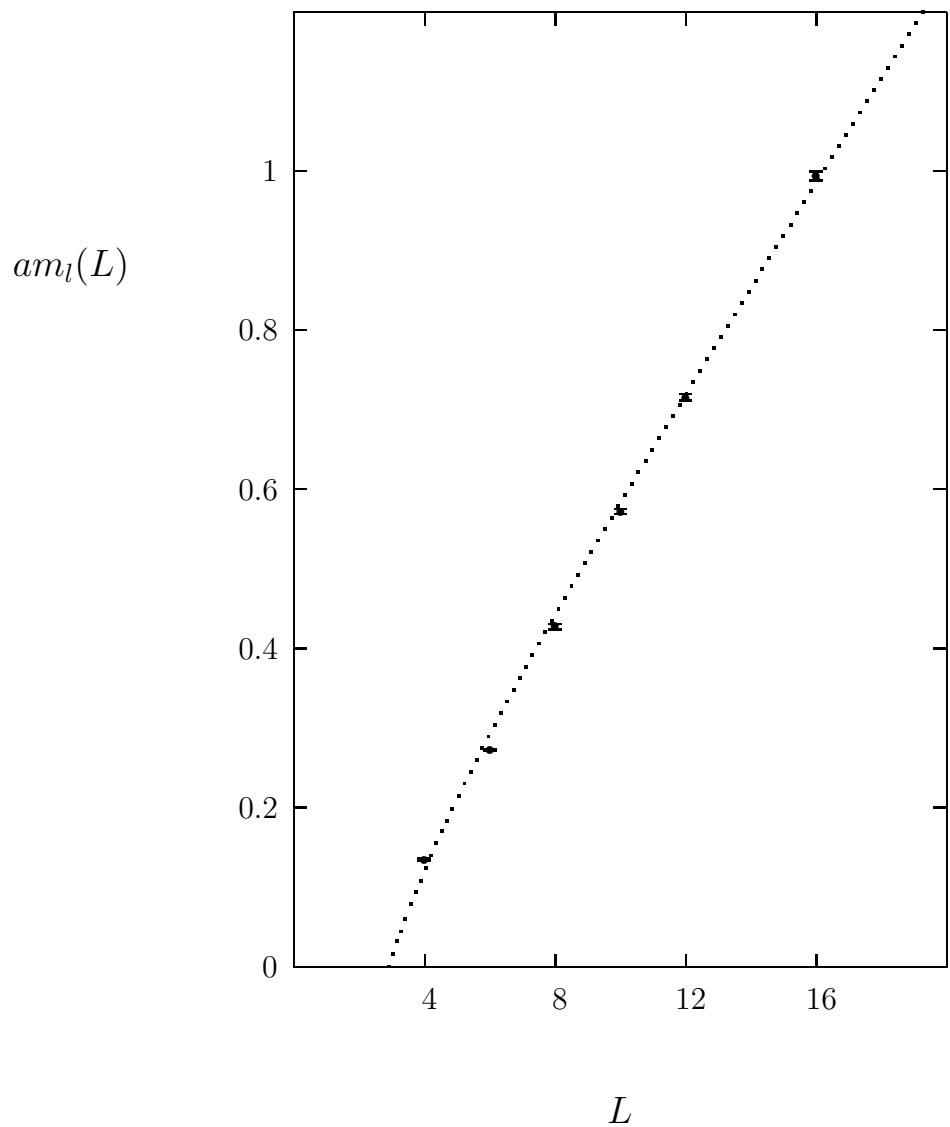

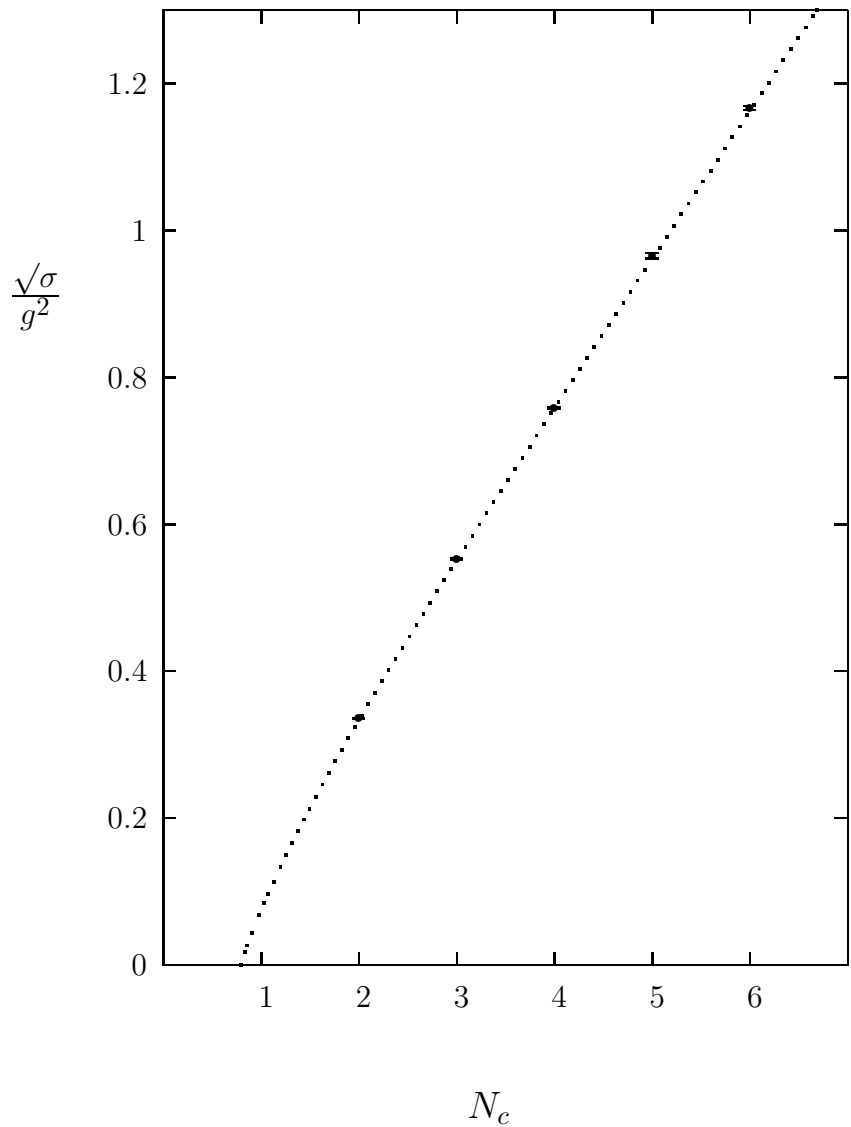
\begin	{figure}[p]
\begin	{center}
\leavevmode
\setlength{\unitlength}{0.240900pt}
\ifx\plotpoint\undefined\newsavebox{\plotpoint}\fi
\sbox{\plotpoint}{\rule[-0.200pt]{0.400pt}{0.400pt}}%
\begin{picture}(1500,1800)(0,0)
\font\gnuplot=cmr10 at 12pt
\gnuplot
\sbox{\plotpoint}{\rule[-0.200pt]{0.400pt}{0.400pt}}%
\put(350.0,250.0){\rule[-0.200pt]{4.818pt}{0.400pt}}
\put(325,250){\makebox(0,0)[r]{\ \ {$0$}}}
\put(1405.0,250.0){\rule[-0.200pt]{4.818pt}{0.400pt}}
\put(350.0,481.0){\rule[-0.200pt]{4.818pt}{0.400pt}}
\put(325,481){\makebox(0,0)[r]{\ \ {$0.2$}}}
\put(1405.0,481.0){\rule[-0.200pt]{4.818pt}{0.400pt}}
\put(350.0,712.0){\rule[-0.200pt]{4.818pt}{0.400pt}}
\put(325,712){\makebox(0,0)[r]{\ \ {$0.4$}}}
\put(1405.0,712.0){\rule[-0.200pt]{4.818pt}{0.400pt}}
\put(350.0,942.0){\rule[-0.200pt]{4.818pt}{0.400pt}}
\put(325,942){\makebox(0,0)[r]{\ \ {$0.6$}}}
\put(1405.0,942.0){\rule[-0.200pt]{4.818pt}{0.400pt}}
\put(350.0,1173.0){\rule[-0.200pt]{4.818pt}{0.400pt}}
\put(325,1173){\makebox(0,0)[r]{\ \ {$0.8$}}}
\put(1405.0,1173.0){\rule[-0.200pt]{4.818pt}{0.400pt}}
\put(350.0,1404.0){\rule[-0.200pt]{4.818pt}{0.400pt}}
\put(325,1404){\makebox(0,0)[r]{\ \ {$1$}}}
\put(1405.0,1404.0){\rule[-0.200pt]{4.818pt}{0.400pt}}
\put(350.0,1635.0){\rule[-0.200pt]{4.818pt}{0.400pt}}
\put(325,1635){\makebox(0,0)[r]{\ \ {$1.2$}}}
\put(1405.0,1635.0){\rule[-0.200pt]{4.818pt}{0.400pt}}
\put(504.0,250.0){\rule[-0.200pt]{0.400pt}{4.818pt}}
\put(504,200){\makebox(0,0){\ {$1$}}}
\put(504.0,1730.0){\rule[-0.200pt]{0.400pt}{4.818pt}}
\put(657.0,250.0){\rule[-0.200pt]{0.400pt}{4.818pt}}
\put(657,200){\makebox(0,0){\ {$2$}}}
\put(657.0,1730.0){\rule[-0.200pt]{0.400pt}{4.818pt}}
\put(811.0,250.0){\rule[-0.200pt]{0.400pt}{4.818pt}}
\put(811,200){\makebox(0,0){\ {$3$}}}
\put(811.0,1730.0){\rule[-0.200pt]{0.400pt}{4.818pt}}
\put(964.0,250.0){\rule[-0.200pt]{0.400pt}{4.818pt}}
\put(964,200){\makebox(0,0){\ {$4$}}}
\put(964.0,1730.0){\rule[-0.200pt]{0.400pt}{4.818pt}}
\put(1118.0,250.0){\rule[-0.200pt]{0.400pt}{4.818pt}}
\put(1118,200){\makebox(0,0){\ {$5$}}}
\put(1118.0,1730.0){\rule[-0.200pt]{0.400pt}{4.818pt}}
\put(1271.0,250.0){\rule[-0.200pt]{0.400pt}{4.818pt}}
\put(1271,200){\makebox(0,0){\ {$6$}}}
\put(1271.0,1730.0){\rule[-0.200pt]{0.400pt}{4.818pt}}
\put(350.0,250.0){\rule[-0.200pt]{258.967pt}{0.400pt}}
\put(1425.0,250.0){\rule[-0.200pt]{0.400pt}{361.350pt}}
\put(350.0,1750.0){\rule[-0.200pt]{258.967pt}{0.400pt}}
\put(150,1300){\makebox(0,0){\Large{${{\surd\sigma} \over g^2}$}}}
\put(862,25){\makebox(0,0){\large{$N_c$}}}
\put(350.0,250.0){\rule[-0.200pt]{0.400pt}{361.350pt}}
\put(657.0,636.0){\rule[-0.200pt]{0.400pt}{0.482pt}}
\put(647.0,636.0){\rule[-0.200pt]{4.818pt}{0.400pt}}
\put(647.0,638.0){\rule[-0.200pt]{4.818pt}{0.400pt}}
\put(811.0,887.0){\rule[-0.200pt]{0.400pt}{0.482pt}}
\put(801.0,887.0){\rule[-0.200pt]{4.818pt}{0.400pt}}
\put(801.0,889.0){\rule[-0.200pt]{4.818pt}{0.400pt}}
\put(964.0,1124.0){\rule[-0.200pt]{0.400pt}{0.723pt}}
\put(954.0,1124.0){\rule[-0.200pt]{4.818pt}{0.400pt}}
\put(954.0,1127.0){\rule[-0.200pt]{4.818pt}{0.400pt}}
\put(1118.0,1360.0){\rule[-0.200pt]{0.400pt}{2.168pt}}
\put(1108.0,1360.0){\rule[-0.200pt]{4.818pt}{0.400pt}}
\put(1108.0,1369.0){\rule[-0.200pt]{4.818pt}{0.400pt}}
\put(1271.0,1593.0){\rule[-0.200pt]{0.400pt}{1.686pt}}
\put(1261.0,1593.0){\rule[-0.200pt]{4.818pt}{0.400pt}}
\put(657,637){\circle*{12}}
\put(811,888){\circle*{12}}
\put(964,1125){\circle*{12}}
\put(1118,1364){\circle*{12}}
\put(1271,1596){\circle*{12}}
\put(1261.0,1600.0){\rule[-0.200pt]{4.818pt}{0.400pt}}
\sbox{\plotpoint}{\rule[-0.500pt]{1.000pt}{1.000pt}}%
\multiput(470,250)(6.563,19.690){2}{\usebox{\plotpoint}}
\multiput(480,280)(7.361,19.406){2}{\usebox{\plotpoint}}
\put(498.69,327.88){\usebox{\plotpoint}}
\put(506.83,346.97){\usebox{\plotpoint}}
\multiput(513,361)(8.648,18.868){2}{\usebox{\plotpoint}}
\put(532.56,403.67){\usebox{\plotpoint}}
\put(541.16,422.56){\usebox{\plotpoint}}
\put(550.13,441.27){\usebox{\plotpoint}}
\put(559.54,459.77){\usebox{\plotpoint}}
\put(569.18,478.15){\usebox{\plotpoint}}
\multiput(578,495)(9.631,18.386){2}{\usebox{\plotpoint}}
\put(598.42,533.12){\usebox{\plotpoint}}
\put(608.75,551.12){\usebox{\plotpoint}}
\put(618.28,569.55){\usebox{\plotpoint}}
\put(628.35,587.69){\usebox{\plotpoint}}
\put(638.75,605.66){\usebox{\plotpoint}}
\put(649.40,623.47){\usebox{\plotpoint}}
\put(659.98,641.32){\usebox{\plotpoint}}
\put(670.60,659.16){\usebox{\plotpoint}}
\put(681.42,676.87){\usebox{\plotpoint}}
\put(691.68,694.90){\usebox{\plotpoint}}
\put(701.87,712.97){\usebox{\plotpoint}}
\put(712.89,730.55){\usebox{\plotpoint}}
\put(723.96,748.11){\usebox{\plotpoint}}
\put(734.78,765.82){\usebox{\plotpoint}}
\put(745.79,783.41){\usebox{\plotpoint}}
\put(756.87,800.96){\usebox{\plotpoint}}
\put(767.56,818.75){\usebox{\plotpoint}}
\put(778.23,836.55){\usebox{\plotpoint}}
\put(789.26,854.13){\usebox{\plotpoint}}
\put(800.54,871.56){\usebox{\plotpoint}}
\put(811.81,888.98){\usebox{\plotpoint}}
\put(822.84,906.56){\usebox{\plotpoint}}
\put(833.90,924.12){\usebox{\plotpoint}}
\put(844.77,941.80){\usebox{\plotpoint}}
\put(855.74,959.42){\usebox{\plotpoint}}
\put(867.02,976.84){\usebox{\plotpoint}}
\put(878.29,994.27){\usebox{\plotpoint}}
\put(889.57,1011.69){\usebox{\plotpoint}}
\put(900.84,1029.12){\usebox{\plotpoint}}
\put(912.46,1046.31){\usebox{\plotpoint}}
\put(923.84,1063.67){\usebox{\plotpoint}}
\put(934.51,1081.47){\usebox{\plotpoint}}
\put(945.68,1098.96){\usebox{\plotpoint}}
\put(956.96,1116.39){\usebox{\plotpoint}}
\put(968.67,1133.52){\usebox{\plotpoint}}
\put(979.96,1150.94){\usebox{\plotpoint}}
\put(991.24,1168.36){\usebox{\plotpoint}}
\put(1002.90,1185.53){\usebox{\plotpoint}}
\put(1013.52,1203.35){\usebox{\plotpoint}}
\put(1024.88,1220.73){\usebox{\plotpoint}}
\put(1036.53,1237.90){\usebox{\plotpoint}}
\put(1047.92,1255.25){\usebox{\plotpoint}}
\put(1059.53,1272.45){\usebox{\plotpoint}}
\put(1070.80,1289.88){\usebox{\plotpoint}}
\put(1081.98,1307.37){\usebox{\plotpoint}}
\put(1093.11,1324.89){\usebox{\plotpoint}}
\put(1104.61,1342.16){\usebox{\plotpoint}}
\put(1116.11,1359.44){\usebox{\plotpoint}}
\put(1127.66,1376.68){\usebox{\plotpoint}}
\put(1139.11,1393.99){\usebox{\plotpoint}}
\put(1150.70,1411.20){\usebox{\plotpoint}}
\put(1161.57,1428.87){\usebox{\plotpoint}}
\put(1173.05,1446.16){\usebox{\plotpoint}}
\put(1184.40,1463.53){\usebox{\plotpoint}}
\put(1196.09,1480.68){\usebox{\plotpoint}}
\put(1207.85,1497.78){\usebox{\plotpoint}}
\put(1219.14,1515.20){\usebox{\plotpoint}}
\put(1230.80,1532.36){\usebox{\plotpoint}}
\put(1241.48,1550.15){\usebox{\plotpoint}}
\put(1253.15,1567.32){\usebox{\plotpoint}}
\put(1264.53,1584.68){\usebox{\plotpoint}}
\put(1276.29,1601.78){\usebox{\plotpoint}}
\put(1287.88,1618.99){\usebox{\plotpoint}}
\put(1299.33,1636.30){\usebox{\plotpoint}}
\put(1310.76,1653.62){\usebox{\plotpoint}}
\put(1321.91,1671.13){\usebox{\plotpoint}}
\put(1333.45,1688.38){\usebox{\plotpoint}}
\put(1345.21,1705.48){\usebox{\plotpoint}}
\put(1356.64,1722.80){\usebox{\plotpoint}}
\put(1368.25,1740.00){\usebox{\plotpoint}}
\put(1375,1750){\usebox{\plotpoint}}
\end{picture}

\end	{center}
\vskip 0.15in
\caption{The value of $\surd\sigma/g^2$ as a function
of $N_c$. The line shows the fit in eqn(\ref{C12}).}
\label{fig_stringN}
\end 	{figure}

\begin	{figure}[p]
\begin	{center}
\leavevmode
\setlength{\unitlength}{0.240900pt}
\ifx\plotpoint\undefined\newsavebox{\plotpoint}\fi
\sbox{\plotpoint}{\rule[-0.200pt]{0.400pt}{0.400pt}}%
\begin{picture}(1500,1800)(0,0)
\font\gnuplot=cmr10 at 12pt
\gnuplot
\sbox{\plotpoint}{\rule[-0.200pt]{0.400pt}{0.400pt}}%
\put(375.0,150.0){\rule[-0.200pt]{4.818pt}{0.400pt}}
\put(350,150){\makebox(0,0)[r]{\ \ {$0$}}}
\put(1405.0,150.0){\rule[-0.200pt]{4.818pt}{0.400pt}}
\put(375.0,264.0){\rule[-0.200pt]{4.818pt}{0.400pt}}
\put(350,264){\makebox(0,0)[r]{\ \ {$2$}}}
\put(1405.0,264.0){\rule[-0.200pt]{4.818pt}{0.400pt}}
\put(375.0,379.0){\rule[-0.200pt]{4.818pt}{0.400pt}}
\put(350,379){\makebox(0,0)[r]{\ \ {$4$}}}
\put(1405.0,379.0){\rule[-0.200pt]{4.818pt}{0.400pt}}
\put(375.0,493.0){\rule[-0.200pt]{4.818pt}{0.400pt}}
\put(350,493){\makebox(0,0)[r]{\ \ {$6$}}}
\put(1405.0,493.0){\rule[-0.200pt]{4.818pt}{0.400pt}}
\put(375.0,607.0){\rule[-0.200pt]{4.818pt}{0.400pt}}
\put(350,607){\makebox(0,0)[r]{\ \ {$8$}}}
\put(1405.0,607.0){\rule[-0.200pt]{4.818pt}{0.400pt}}
\put(375.0,721.0){\rule[-0.200pt]{4.818pt}{0.400pt}}
\put(350,721){\makebox(0,0)[r]{\ \ {$10$}}}
\put(1405.0,721.0){\rule[-0.200pt]{4.818pt}{0.400pt}}
\put(375.0,836.0){\rule[-0.200pt]{4.818pt}{0.400pt}}
\put(350,836){\makebox(0,0)[r]{\ \ {$12$}}}
\put(1405.0,836.0){\rule[-0.200pt]{4.818pt}{0.400pt}}
\put(375.0,950.0){\rule[-0.200pt]{4.818pt}{0.400pt}}
\put(350,950){\makebox(0,0)[r]{\ \ {$14$}}}
\put(1405.0,950.0){\rule[-0.200pt]{4.818pt}{0.400pt}}
\put(375.0,1064.0){\rule[-0.200pt]{4.818pt}{0.400pt}}
\put(350,1064){\makebox(0,0)[r]{\ \ {$16$}}}
\put(1405.0,1064.0){\rule[-0.200pt]{4.818pt}{0.400pt}}
\put(375.0,1179.0){\rule[-0.200pt]{4.818pt}{0.400pt}}
\put(350,1179){\makebox(0,0)[r]{\ \ {$18$}}}
\put(1405.0,1179.0){\rule[-0.200pt]{4.818pt}{0.400pt}}
\put(375.0,1293.0){\rule[-0.200pt]{4.818pt}{0.400pt}}
\put(350,1293){\makebox(0,0)[r]{\ \ {$20$}}}
\put(1405.0,1293.0){\rule[-0.200pt]{4.818pt}{0.400pt}}
\put(375.0,1407.0){\rule[-0.200pt]{4.818pt}{0.400pt}}
\put(350,1407){\makebox(0,0)[r]{\ \ {$22$}}}
\put(1405.0,1407.0){\rule[-0.200pt]{4.818pt}{0.400pt}}
\put(375.0,1521.0){\rule[-0.200pt]{4.818pt}{0.400pt}}
\put(350,1521){\makebox(0,0)[r]{\ \ {$24$}}}
\put(1405.0,1521.0){\rule[-0.200pt]{4.818pt}{0.400pt}}
\put(463.0,150.0){\rule[-0.200pt]{0.400pt}{4.818pt}}
\put(463,100){\makebox(0,0){ {$1$}}}
\put(463.0,1730.0){\rule[-0.200pt]{0.400pt}{4.818pt}}
\put(681.0,150.0){\rule[-0.200pt]{0.400pt}{4.818pt}}
\put(681,100){\makebox(0,0){ {$1.5$}}}
\put(681.0,1730.0){\rule[-0.200pt]{0.400pt}{4.818pt}}
\put(900.0,150.0){\rule[-0.200pt]{0.400pt}{4.818pt}}
\put(900,100){\makebox(0,0){ {$2$}}}
\put(900.0,1730.0){\rule[-0.200pt]{0.400pt}{4.818pt}}
\put(1119.0,150.0){\rule[-0.200pt]{0.400pt}{4.818pt}}
\put(1119,100){\makebox(0,0){ {$2.5$}}}
\put(1119.0,1730.0){\rule[-0.200pt]{0.400pt}{4.818pt}}
\put(1338.0,150.0){\rule[-0.200pt]{0.400pt}{4.818pt}}
\put(1338,100){\makebox(0,0){ {$3$}}}
\put(1338.0,1730.0){\rule[-0.200pt]{0.400pt}{4.818pt}}
\put(375.0,150.0){\rule[-0.200pt]{252.945pt}{0.400pt}}
\put(1425.0,150.0){\rule[-0.200pt]{0.400pt}{385.440pt}}
\put(375.0,1750.0){\rule[-0.200pt]{252.945pt}{0.400pt}}
\put(50,1000){\makebox(0,0){\Large{${{\chi^2}\over{n_{df}}}$}}}
\put(900,25){\makebox(0,0){\Large{$\alpha$}}}
\put(375.0,150.0){\rule[-0.200pt]{0.400pt}{385.440pt}}
\put(463,1550){\circle*{12}}
\put(471,1489){\circle*{12}}
\put(480,1430){\circle*{12}}
\put(489,1372){\circle*{12}}
\put(498,1315){\circle*{12}}
\put(506,1260){\circle*{12}}
\put(515,1207){\circle*{12}}
\put(524,1156){\circle*{12}}
\put(532,1105){\circle*{12}}
\put(541,1057){\circle*{12}}
\put(550,1009){\circle*{12}}
\put(559,964){\circle*{12}}
\put(568,919){\circle*{12}}
\put(576,877){\circle*{12}}
\put(585,835){\circle*{12}}
\put(594,795){\circle*{12}}
\put(603,757){\circle*{12}}
\put(611,720){\circle*{12}}
\put(620,684){\circle*{12}}
\put(629,650){\circle*{12}}
\put(637,617){\circle*{12}}
\put(646,585){\circle*{12}}
\put(655,555){\circle*{12}}
\put(664,526){\circle*{12}}
\put(673,499){\circle*{12}}
\put(681,472){\circle*{12}}
\put(690,447){\circle*{12}}
\put(699,424){\circle*{12}}
\put(708,401){\circle*{12}}
\put(716,380){\circle*{12}}
\put(725,360){\circle*{12}}
\put(734,341){\circle*{12}}
\put(742,323){\circle*{12}}
\put(751,307){\circle*{12}}
\put(760,291){\circle*{12}}
\put(769,277){\circle*{12}}
\put(778,264){\circle*{12}}
\put(786,252){\circle*{12}}
\put(795,241){\circle*{12}}
\put(804,232){\circle*{12}}
\put(813,223){\circle*{12}}
\put(821,215){\circle*{12}}
\put(830,209){\circle*{12}}
\put(839,203){\circle*{12}}
\put(847,199){\circle*{12}}
\put(856,195){\circle*{12}}
\put(865,193){\circle*{12}}
\put(874,191){\circle*{12}}
\put(882,190){\circle*{12}}
\put(891,191){\circle*{12}}
\put(900,192){\circle*{12}}
\put(909,194){\circle*{12}}
\put(917,197){\circle*{12}}
\put(926,201){\circle*{12}}
\put(935,206){\circle*{12}}
\put(944,211){\circle*{12}}
\put(953,217){\circle*{12}}
\put(961,225){\circle*{12}}
\put(970,233){\circle*{12}}
\put(979,241){\circle*{12}}
\put(988,251){\circle*{12}}
\put(996,261){\circle*{12}}
\put(1005,272){\circle*{12}}
\put(1014,284){\circle*{12}}
\put(1022,297){\circle*{12}}
\put(1031,310){\circle*{12}}
\put(1040,324){\circle*{12}}
\put(1049,338){\circle*{12}}
\put(1058,353){\circle*{12}}
\put(1066,369){\circle*{12}}
\put(1075,386){\circle*{12}}
\put(1084,403){\circle*{12}}
\put(1093,420){\circle*{12}}
\put(1101,439){\circle*{12}}
\put(1110,458){\circle*{12}}
\put(1119,477){\circle*{12}}
\put(1128,497){\circle*{12}}
\put(1136,517){\circle*{12}}
\put(1145,538){\circle*{12}}
\put(1154,560){\circle*{12}}
\put(1163,582){\circle*{12}}
\put(1171,605){\circle*{12}}
\put(1180,628){\circle*{12}}
\put(1189,651){\circle*{12}}
\put(1198,675){\circle*{12}}
\put(1206,700){\circle*{12}}
\put(1215,724){\circle*{12}}
\put(1224,750){\circle*{12}}
\put(1232,775){\circle*{12}}
\put(1241,801){\circle*{12}}
\put(1250,828){\circle*{12}}
\put(1259,855){\circle*{12}}
\put(1268,882){\circle*{12}}
\put(1276,909){\circle*{12}}
\put(1285,937){\circle*{12}}
\put(1294,966){\circle*{12}}
\put(1303,994){\circle*{12}}
\put(1311,1023){\circle*{12}}
\put(1320,1052){\circle*{12}}
\put(1329,1082){\circle*{12}}
\put(1338,1112){\circle*{12}}
\end{picture}

\end	{center}
\vskip 0.15in
\caption{The $\chi^2$ per degree of freedom against
the power, $\alpha$, of the leading large-$N_c$ correction
when fitting $\surd\sigma/g^2N_c$ to eqn(\ref{C4}).}
\label{fig_corrN}
\end 	{figure}
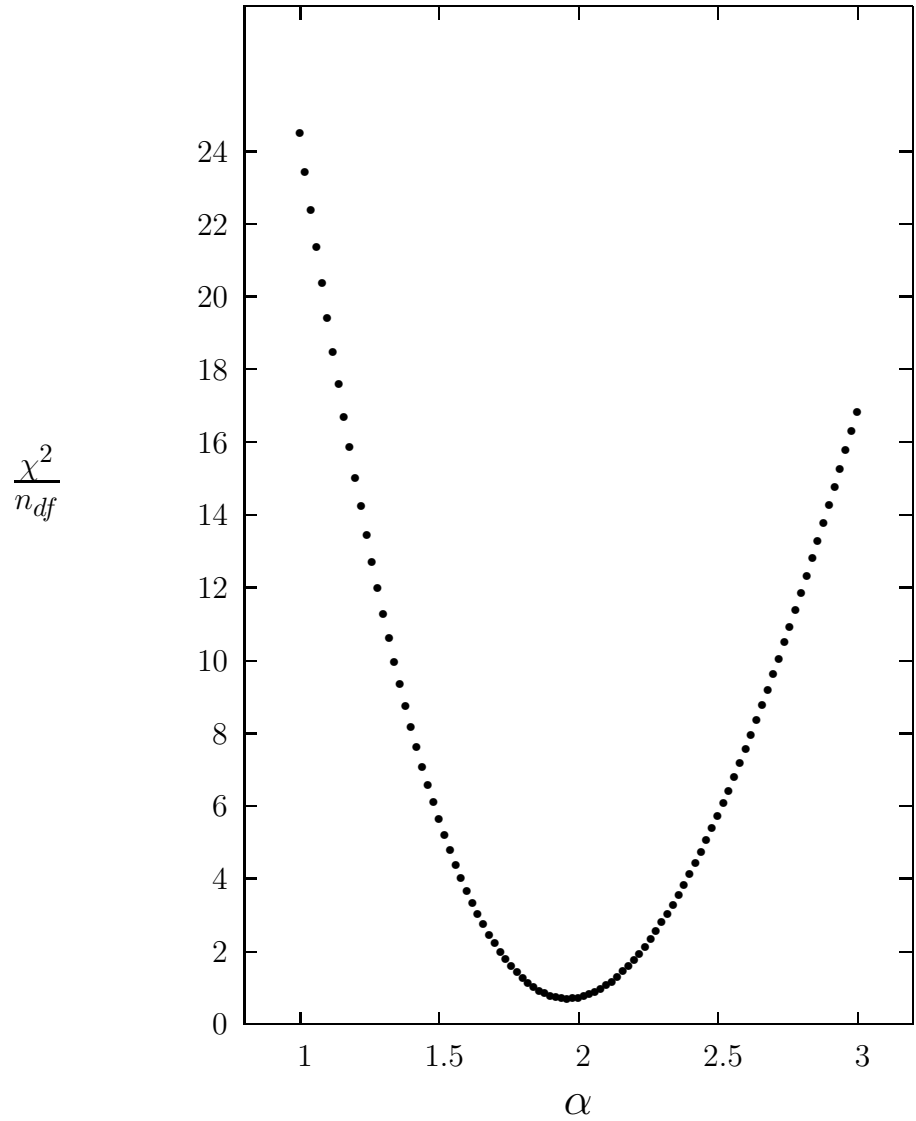

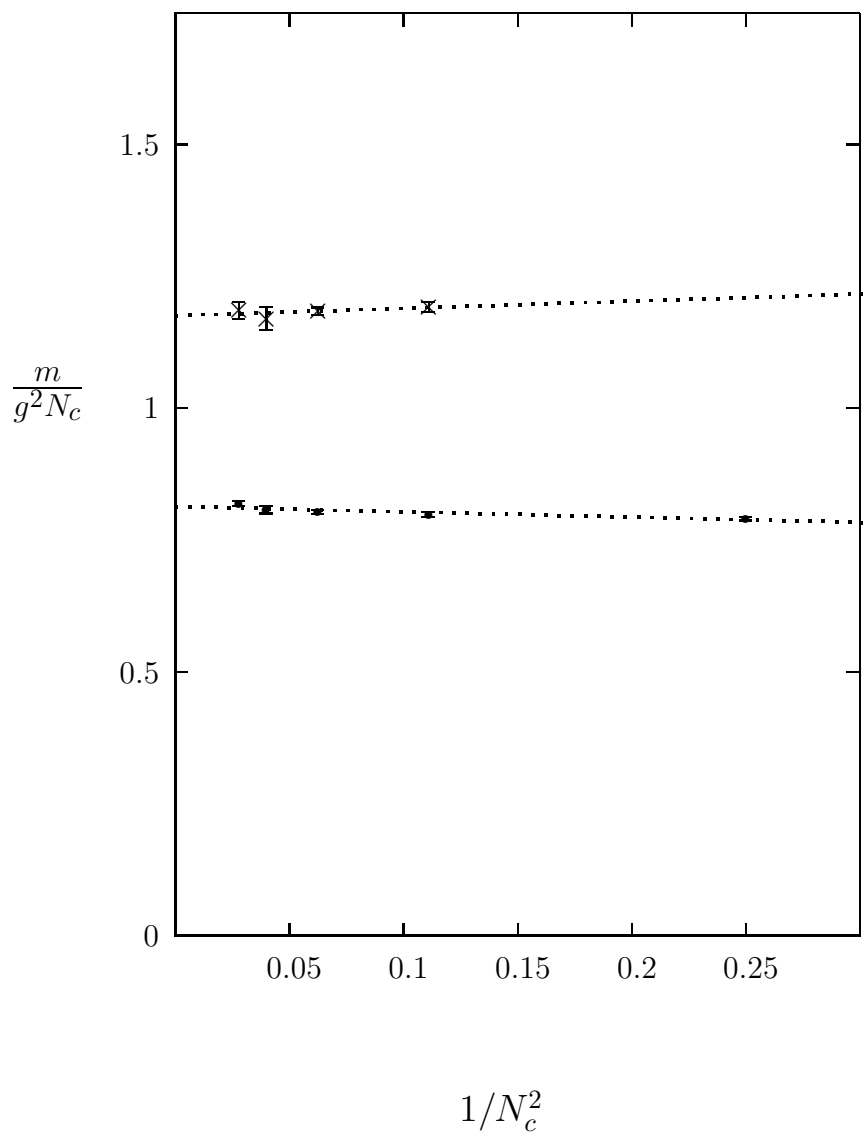
\begin	{figure}[p]
\begin	{center}
\leavevmode
\setlength{\unitlength}{0.240900pt}
\ifx\plotpoint\undefined\newsavebox{\plotpoint}\fi
\sbox{\plotpoint}{\rule[-0.200pt]{0.400pt}{0.400pt}}%
\begin{picture}(1500,1800)(0,0)
\font\gnuplot=cmr10 at 12pt
\gnuplot
\sbox{\plotpoint}{\rule[-0.200pt]{0.400pt}{0.400pt}}%
\put(350.0,300.0){\rule[-0.200pt]{4.818pt}{0.400pt}}
\put(325,300){\makebox(0,0)[r]{\ \ {$0$}}}
\put(1405.0,300.0){\rule[-0.200pt]{4.818pt}{0.400pt}}
\put(350.0,714.0){\rule[-0.200pt]{4.818pt}{0.400pt}}
\put(325,714){\makebox(0,0)[r]{\ \ {$0.5$}}}
\put(1405.0,714.0){\rule[-0.200pt]{4.818pt}{0.400pt}}
\put(350.0,1129.0){\rule[-0.200pt]{4.818pt}{0.400pt}}
\put(325,1129){\makebox(0,0)[r]{\ \ {$1$}}}
\put(1405.0,1129.0){\rule[-0.200pt]{4.818pt}{0.400pt}}
\put(350.0,1543.0){\rule[-0.200pt]{4.818pt}{0.400pt}}
\put(325,1543){\makebox(0,0)[r]{\ \ {$1.5$}}}
\put(1405.0,1543.0){\rule[-0.200pt]{4.818pt}{0.400pt}}
\put(529.0,300.0){\rule[-0.200pt]{0.400pt}{4.818pt}}
\put(529,250){\makebox(0,0){\ {$0.05$}}}
\put(529.0,1730.0){\rule[-0.200pt]{0.400pt}{4.818pt}}
\put(708.0,300.0){\rule[-0.200pt]{0.400pt}{4.818pt}}
\put(708,250){\makebox(0,0){\ {$0.1$}}}
\put(708.0,1730.0){\rule[-0.200pt]{0.400pt}{4.818pt}}
\put(888.0,300.0){\rule[-0.200pt]{0.400pt}{4.818pt}}
\put(888,250){\makebox(0,0){\ {$0.15$}}}
\put(888.0,1730.0){\rule[-0.200pt]{0.400pt}{4.818pt}}
\put(1067.0,300.0){\rule[-0.200pt]{0.400pt}{4.818pt}}
\put(1067,250){\makebox(0,0){\ {$0.2$}}}
\put(1067.0,1730.0){\rule[-0.200pt]{0.400pt}{4.818pt}}
\put(1246.0,300.0){\rule[-0.200pt]{0.400pt}{4.818pt}}
\put(1246,250){\makebox(0,0){\ {$0.25$}}}
\put(1246.0,1730.0){\rule[-0.200pt]{0.400pt}{4.818pt}}
\put(350.0,300.0){\rule[-0.200pt]{258.967pt}{0.400pt}}
\put(1425.0,300.0){\rule[-0.200pt]{0.400pt}{349.305pt}}
\put(350.0,1750.0){\rule[-0.200pt]{258.967pt}{0.400pt}}
\put(150,1150){\makebox(0,0){\Large{${m \over {g^2 N_c}}$}}}
\put(862,25){\makebox(0,0){\large{$1/N^2_c$}}}
\put(350.0,300.0){\rule[-0.200pt]{0.400pt}{349.305pt}}
\put(1246.0,952.0){\rule[-0.200pt]{0.400pt}{1.445pt}}
\put(1236.0,952.0){\rule[-0.200pt]{4.818pt}{0.400pt}}
\put(1236.0,958.0){\rule[-0.200pt]{4.818pt}{0.400pt}}
\put(748.0,958.0){\rule[-0.200pt]{0.400pt}{1.686pt}}
\put(738.0,958.0){\rule[-0.200pt]{4.818pt}{0.400pt}}
\put(738.0,965.0){\rule[-0.200pt]{4.818pt}{0.400pt}}
\put(574.0,962.0){\rule[-0.200pt]{0.400pt}{1.686pt}}
\put(564.0,962.0){\rule[-0.200pt]{4.818pt}{0.400pt}}
\put(564.0,969.0){\rule[-0.200pt]{4.818pt}{0.400pt}}
\put(493.0,963.0){\rule[-0.200pt]{0.400pt}{2.891pt}}
\put(483.0,963.0){\rule[-0.200pt]{4.818pt}{0.400pt}}
\put(483.0,975.0){\rule[-0.200pt]{4.818pt}{0.400pt}}
\put(450.0,975.0){\rule[-0.200pt]{0.400pt}{1.927pt}}
\put(440.0,975.0){\rule[-0.200pt]{4.818pt}{0.400pt}}
\put(1246,955){\circle*{12}}
\put(748,961){\circle*{12}}
\put(574,966){\circle*{12}}
\put(493,969){\circle*{12}}
\put(450,979){\circle*{12}}
\put(440.0,983.0){\rule[-0.200pt]{4.818pt}{0.400pt}}
\put(748.0,1280.0){\rule[-0.200pt]{0.400pt}{3.613pt}}
\put(738.0,1280.0){\rule[-0.200pt]{4.818pt}{0.400pt}}
\put(738.0,1295.0){\rule[-0.200pt]{4.818pt}{0.400pt}}
\put(574.0,1275.0){\rule[-0.200pt]{0.400pt}{2.891pt}}
\put(564.0,1275.0){\rule[-0.200pt]{4.818pt}{0.400pt}}
\put(564.0,1287.0){\rule[-0.200pt]{4.818pt}{0.400pt}}
\put(493.0,1251.0){\rule[-0.200pt]{0.400pt}{8.913pt}}
\put(483.0,1251.0){\rule[-0.200pt]{4.818pt}{0.400pt}}
\put(483.0,1288.0){\rule[-0.200pt]{4.818pt}{0.400pt}}
\put(450.0,1269.0){\rule[-0.200pt]{0.400pt}{6.504pt}}
\put(440.0,1269.0){\rule[-0.200pt]{4.818pt}{0.400pt}}
\put(748,1288){\makebox(0,0){$\times$}}
\put(574,1281){\makebox(0,0){$\times$}}
\put(493,1269){\makebox(0,0){$\times$}}
\put(450,1283){\makebox(0,0){$\times$}}
\put(440.0,1296.0){\rule[-0.200pt]{4.818pt}{0.400pt}}
\sbox{\plotpoint}{\rule[-0.500pt]{1.000pt}{1.000pt}}%
\put(350,974){\usebox{\plotpoint}}
\put(350.00,974.00){\usebox{\plotpoint}}
\put(370.76,974.00){\usebox{\plotpoint}}
\put(391.47,973.15){\usebox{\plotpoint}}
\put(412.22,973.00){\usebox{\plotpoint}}
\put(432.94,972.37){\usebox{\plotpoint}}
\put(453.68,972.00){\usebox{\plotpoint}}
\put(474.42,971.51){\usebox{\plotpoint}}
\put(495.15,971.00){\usebox{\plotpoint}}
\put(515.86,970.00){\usebox{\plotpoint}}
\put(536.61,970.00){\usebox{\plotpoint}}
\put(557.32,969.00){\usebox{\plotpoint}}
\put(578.08,969.00){\usebox{\plotpoint}}
\put(598.79,968.11){\usebox{\plotpoint}}
\put(619.54,968.00){\usebox{\plotpoint}}
\put(640.27,967.25){\usebox{\plotpoint}}
\put(661.01,967.00){\usebox{\plotpoint}}
\put(681.74,966.48){\usebox{\plotpoint}}
\put(702.48,966.00){\usebox{\plotpoint}}
\put(723.21,965.62){\usebox{\plotpoint}}
\put(743.94,965.00){\usebox{\plotpoint}}
\put(764.69,964.83){\usebox{\plotpoint}}
\put(785.40,964.00){\usebox{\plotpoint}}
\put(806.16,963.99){\usebox{\plotpoint}}
\put(826.87,963.00){\usebox{\plotpoint}}
\put(847.62,963.00){\usebox{\plotpoint}}
\put(868.33,962.00){\usebox{\plotpoint}}
\put(889.09,962.00){\usebox{\plotpoint}}
\put(909.80,961.00){\usebox{\plotpoint}}
\put(930.56,961.00){\usebox{\plotpoint}}
\put(951.27,960.00){\usebox{\plotpoint}}
\put(972.02,960.00){\usebox{\plotpoint}}
\put(992.73,959.00){\usebox{\plotpoint}}
\put(1013.49,959.00){\usebox{\plotpoint}}
\put(1034.20,958.00){\usebox{\plotpoint}}
\put(1054.95,958.00){\usebox{\plotpoint}}
\put(1075.67,957.21){\usebox{\plotpoint}}
\put(1096.42,957.00){\usebox{\plotpoint}}
\put(1117.14,956.35){\usebox{\plotpoint}}
\put(1137.88,956.00){\usebox{\plotpoint}}
\put(1158.62,955.54){\usebox{\plotpoint}}
\put(1179.34,955.00){\usebox{\plotpoint}}
\put(1200.09,954.72){\usebox{\plotpoint}}
\put(1220.81,954.00){\usebox{\plotpoint}}
\put(1241.56,953.86){\usebox{\plotpoint}}
\put(1262.28,953.00){\usebox{\plotpoint}}
\put(1283.03,953.00){\usebox{\plotpoint}}
\put(1303.74,952.00){\usebox{\plotpoint}}
\put(1324.50,952.00){\usebox{\plotpoint}}
\put(1345.21,951.00){\usebox{\plotpoint}}
\put(1365.96,951.00){\usebox{\plotpoint}}
\put(1386.67,950.00){\usebox{\plotpoint}}
\put(1407.43,950.00){\usebox{\plotpoint}}
\put(1425,949){\usebox{\plotpoint}}
\put(350,1274){\usebox{\plotpoint}}
\put(350.00,1274.00){\usebox{\plotpoint}}
\put(370.71,1275.00){\usebox{\plotpoint}}
\put(391.42,1275.84){\usebox{\plotpoint}}
\put(412.17,1276.00){\usebox{\plotpoint}}
\put(432.88,1277.00){\usebox{\plotpoint}}
\put(453.61,1277.51){\usebox{\plotpoint}}
\put(474.33,1278.48){\usebox{\plotpoint}}
\put(495.06,1279.00){\usebox{\plotpoint}}
\put(515.77,1280.00){\usebox{\plotpoint}}
\put(536.52,1280.15){\usebox{\plotpoint}}
\put(557.23,1281.00){\usebox{\plotpoint}}
\put(577.94,1281.99){\usebox{\plotpoint}}
\put(598.69,1282.00){\usebox{\plotpoint}}
\put(619.40,1283.00){\usebox{\plotpoint}}
\put(640.13,1283.74){\usebox{\plotpoint}}
\put(660.87,1284.00){\usebox{\plotpoint}}
\put(681.58,1285.00){\usebox{\plotpoint}}
\put(702.31,1285.48){\usebox{\plotpoint}}
\put(723.05,1286.00){\usebox{\plotpoint}}
\put(743.76,1287.00){\usebox{\plotpoint}}
\put(764.50,1287.15){\usebox{\plotpoint}}
\put(785.22,1288.00){\usebox{\plotpoint}}
\put(805.93,1288.99){\usebox{\plotpoint}}
\put(826.68,1289.00){\usebox{\plotpoint}}
\put(847.39,1290.00){\usebox{\plotpoint}}
\put(868.11,1290.74){\usebox{\plotpoint}}
\put(888.86,1291.00){\usebox{\plotpoint}}
\put(909.57,1292.00){\usebox{\plotpoint}}
\put(930.28,1293.00){\usebox{\plotpoint}}
\put(951.02,1293.37){\usebox{\plotpoint}}
\put(971.74,1294.00){\usebox{\plotpoint}}
\put(992.45,1295.00){\usebox{\plotpoint}}
\put(1013.20,1295.11){\usebox{\plotpoint}}
\put(1033.92,1296.00){\usebox{\plotpoint}}
\put(1054.64,1296.88){\usebox{\plotpoint}}
\put(1075.39,1297.00){\usebox{\plotpoint}}
\put(1096.09,1298.00){\usebox{\plotpoint}}
\put(1116.82,1298.62){\usebox{\plotpoint}}
\put(1137.56,1299.00){\usebox{\plotpoint}}
\put(1158.27,1300.00){\usebox{\plotpoint}}
\put(1179.01,1300.36){\usebox{\plotpoint}}
\put(1199.73,1301.00){\usebox{\plotpoint}}
\put(1220.44,1302.00){\usebox{\plotpoint}}
\put(1241.19,1302.11){\usebox{\plotpoint}}
\put(1261.91,1303.00){\usebox{\plotpoint}}
\put(1282.62,1303.88){\usebox{\plotpoint}}
\put(1303.37,1304.00){\usebox{\plotpoint}}
\put(1324.08,1305.00){\usebox{\plotpoint}}
\put(1344.79,1306.00){\usebox{\plotpoint}}
\put(1365.52,1306.50){\usebox{\plotpoint}}
\put(1386.26,1307.00){\usebox{\plotpoint}}
\put(1406.97,1308.00){\usebox{\plotpoint}}
\put(1425,1308){\usebox{\plotpoint}}
\end{picture}

\end	{center}
\vskip 0.15in
\caption{$0^{++}$($\bullet$) and $0^{--}$($\times$) glueball 
masses in units of $g^2N_c$, plotted against $1/N_c^2$. The best 
linear extrapolations to the $N_c = \infty$ limit are shown.}
\label{fig_msuN}
\end 	{figure}

\end{document}